\documentclass[conference]{IEEEtran}
\IEEEoverridecommandlockouts

\usepackage{cite}
\usepackage{amsmath,amssymb,amsfonts}

\usepackage{xcolor}

\usepackage{tikz}
\usepackage{algorithm}
\usepackage[noend]{algpseudocode} 
\usepackage{textcomp}
\usepackage{booktabs}
\usepackage{multirow}
\usepackage{colortbl}
\usepackage{balance}
\usepackage[inline]{enumitem}
\usepackage{listings}
\usepackage{url}
\usepackage{hyperref}       
\usetikzlibrary{arrows,shapes}

\def\BibTeX{{\rm B\kern-.05em{\sc i\kern-.025em b}\kern-.08em
    T\kern-.1667em\lower.7ex\hbox{E}\kern-.125emX}}
\begin{document}

\title{Approximate Query Processing via Tuple Bubbles}

\author{\IEEEauthorblockN{Damjan Gjurovski}
\IEEEauthorblockA{
\textit{TU Kaiserslautern (TUK)}\\
Kaiserslautern, Germany \\
gjurovski@cs.uni-kl.de}
\and
\IEEEauthorblockN{Sebastian Michel}
\IEEEauthorblockA{
\textit{TU Kaiserslautern (TUK)}\\
Kaiserslautern, Germany \\
michel@cs.uni-kl.de}
}

\pagenumbering{arabic}
\pagestyle{plain}

\maketitle

\begin{abstract}
We propose a versatile approach to lightweight, approximate query processing by creating compact but tunably precise representations of larger quantities of original tuples, coined bubbles.  Instead of working with tables of tuples, the query processing then operates on bubbles but leaves the traditional query processing paradigms conceptually applicable. We believe this is a natural and viable approach to render approximate query processing feasible for large data in disaggregated cloud settings or in resource-limited scenarios.  Bubbles are tunable regarding the compactness of enclosed tuples as well as the granularity of statistics and the way they are instantiated. For improved accuracy, we put forward a first working solution that represents bubbles via Bayesian networks, per table, or along foreign-key joins. To underpin the viability of the approach, we report on an experimental evaluation considering the state-of-the-art competitors, where we show clear benefits when assessing the estimation accuracy, execution time, and required disk space. 
\end{abstract}


\section{Introduction}
\label{sec:introduction}

For data-intensive tasks under latency or resource constraints, approaches on approximate query processing (AQP) are viable solutions 
if exact answers are not imperative.  This particularly applies to areas such as  data exploration and visualization where the approximate answers can be used for discovering trends~\cite{DBLP:journals/tvcg/ZgraggenGCFK17} or identifying parts of the data that need further processing through interactive analysis~\cite{peng2018aqp++,DBLP:journals/pvldb/GalakatosCZBK17}.

Recent approaches for approximate query processing make use of pre-computed aggregates or sampling, or a combination of both, for a more accurate question answering~\cite{liang2021combining,peng2018aqp++,park2018verdictdb}. Pure \textit{sampling-based} approaches usually create samples that cannot capture highly selective queries, and normally they produce inadequate results for such queries. To avoid this, most sampling-based solutions make use of workload information~\cite{DBLP:conf/eurosys/AgarwalMPMMS13,DBLP:conf/sigmod/DingHCC016,DBLP:conf/sigmod/ChaudhuriDN01}. Approaches that are based on pre-computed aggregates~\cite{DBLP:journals/datamine/GrayCBLRVPP97,DBLP:journals/pvldb/GalakatosCZBK17} store the answers to some aggregation queries and use them for improving the query performance. Recently proposed AQP approaches that also combine pre-computation with sampling are limited to the queries that they can answer since they either cannot answer any join queries~\cite{liang2021combining,peng2018aqp++} or they require the creation of large samples that drastically affects the execution time~\cite{park2018verdictdb}. 

Our proposed approach lifts traditional processing of relational tuples  to processing so-called \textbf{tuple bubbles}, or briefly put bubbles. Bubbles represent groups of tuples from tables in a compact way. In other words,  bubbles are summaries of chunks of tuples, typically from one partitioned relation, but possibly also from pre-computed results for frequent sub queries.
 When a user query arrives, an approximate answer is derived based solely on the bubbles. 
We expect that there will be orders of magnitude less bubbles than tuples, which can vastly accelerate the query processing.
Further, in distributed environments, especially in so-called disaggregated settings where data shipping is mandatory, bubbles can not only deliver approximate query results in a bandwidth-saving manner, but can enable reliable dataset discovery in data lakes.

In this first work, we focus on the representation of bubbles 
and how to perform query processing over them. 
To organize the tuples into bubbles we employ a plain horizontal partitioning schema. 
However, the assignment of tuples to bubbles is a challenge on its own that requires detailed analysis. If done right, it is expected to greatly boost the ability of statistics to represent tuples.

While there are multiple ways to succinctly summarize tuples contained in a bubble, like sampling or  histograms, we propose the usage of \textbf{Bayesian networks}. Even though Bayesian networks have been successfully applied for the task of answering count queries, i.e., cardinality estimation~\cite{DBLP:journals/pvldb/TzoumasDJ11,DBLP:conf/dasfaa/HalfordSM19,DBLP:journals/tlsdkcs/HalfordSM20,DBLP:journals/corr/abs-2012-14743}, to the best of our knowledge, this is the first approach that employs them for AQP. Through the use of Bayesian networks we produce estimates in a timely manner while efficiently representing   dependencies between correlated attributes. Additionally, they  allow the connection of estimates from different tuple bubbles, which is crucial for answering aggregation queries that involve joins. 


\subsection{Problem Formulation}
\label{sec:problem_formulation}
We consider a database consisting of relations $R_i$, each having a set of attributes $A_1, ..., A_n$. 
For each relation, we want to assign tuples into $k$ disjoint partitions, called tuple bubbles $TB_1, ..., TB_k$.
Instead of storing the complete tuples in a bubble, they need to be represented through statistics. The created per-bubble statistics are then used for approximating the answer of \textbf{aggregation queries}, i.e., \texttt{SUM}, \texttt{AVG}, \texttt{MIN}, \texttt{MAX}, and \texttt{COUNT}, involving an arbitrary number of \textbf{equality joins} and \textbf{equality or range predicates}.

\subsection{Contributions and Outline}
\label{ssec:contributions_and_outline}
The main contributions of this paper are:
\begin{itemize}
	\item We propose an approach for approximate query processing that performs the processing over groups of tuples, i.e., tuple bubbles, instead of individual tuples (Section~\ref{sec:approach}). 
	\item For efficient but accurate computation, we represent the bubbles through statistics using Bayesian networks (Section~\ref{ssec:tuple_bubbles}) and combine the results for the final estimate (Section~\ref{ssec:aqp_over_tuple_bubbles}) considering queries with arbitrary number of joins and predicates (Section~\ref{sec:aqp_over_tuple_bubbles}).
	\item We detail the encountered challenges and possible solutions (Section~\ref{sec:challenges_future_outlook}), and report the promising results of our evaluation (Section~\ref{sec:experiments}).

\end{itemize}


\section{Related Work}
\label{sec:relatedwork}

Sampling-based approaches for AQP have been widely researched, dating back several decades~\cite{DBLP:conf/sigmod/ChaudhuriDK17,DBLP:journals/ftdb/CormodeGHJ12}. However, these approaches come with limitations that are well known. As one solution, many sampling based approaches use information for the query workloads to produce better results~\cite{DBLP:conf/eurosys/AgarwalMPMMS13,DBLP:conf/sigmod/ChaudhuriDN01,DBLP:conf/sigmod/DingHCC016}. On the other spectrum, there are many approaches that create samples online~\cite{DBLP:conf/sigmod/KandulaSVOGCD16,DBLP:journals/tods/LiWYZ19,DBLP:conf/sigmod/WuOT10}. Wander join~\cite{DBLP:journals/tods/LiWYZ19} creates estimates by modeling the tables as a join graph and performs random walks through the graph. VerdictDB~\cite{park2018verdictdb} is an AQP system that supports approximate query processing of general ad-hoc queries. It works over so called scrambles which represent samples from the original data. It uses only the created samples for producing result estimates. Both Wander join and VerdictDB are capable of answering queries with joins. We will consider both in our experimental evaluation.

In addition, there exist various approaches that use precomputed aggregates or a combination of precomputed aggregates and sampling for approximate query processing~\cite{liang2021combining,peng2018aqp++,DBLP:journals/datamine/GrayCBLRVPP97,DBLP:journals/pvldb/GalakatosCZBK17}. Liang et al.~\cite{liang2021combining} combine precomputed aggregates with stratified sampling for producing query estimates. They propose an indexing structure (PASS) that represents a hierarchal partitioning of the dataset such that for every partition in the tree they store the \texttt{MIN}, \texttt{MAX}, \texttt{SUM}, and \texttt{COUNT}. The leaf nodes store a uniform sample of the data from the respective partition. 
However, the approach is limited to only answering aggregation queries without any joins. AQP++~\cite{peng2018aqp++} is a similar approach to PASS which precomputes the results to numerous aggregation queries. Additionally, it determines a query subsumption relationships to match a new query to an existing one and then uses uniform samples to approximate the gap. AQP++ can also only answer queries that do not have any joins. We consider both in our evaluation. 

There also exist approaches that apply machine learning for approximate query processing~\cite{DBLP:journals/pvldb/YangLKWDCAHKS19,DBLP:journals/pvldb/HilprechtSKMKB20,DBLP:journals/pvldb/WalenzSRY19}. DeepDB~\cite{DBLP:journals/pvldb/HilprechtSKMKB20} learns directly over the data by using relational sum product networks to efficiently capture the database characteristics. Naru~\cite{DBLP:journals/pvldb/YangLKWDCAHKS19} uses deep autoregressive models and trains directly over the data for selectivity estimation. The underlying models of both approaches can be considered as a replacement of Bayesian networks in our proposed approach. 

It is important to point out that although our approach aims at estimating results for aggregation queries, we want to accomplish this by moving the computation from tuples to groups or partitions of them, or so-called tuple bubbles. In the future, many approaches that support joins can be potentially used for storing the per tuple bubbles statistics. 


\begin{figure*}[!t]
	\center
	\includegraphics[width=1.\linewidth]{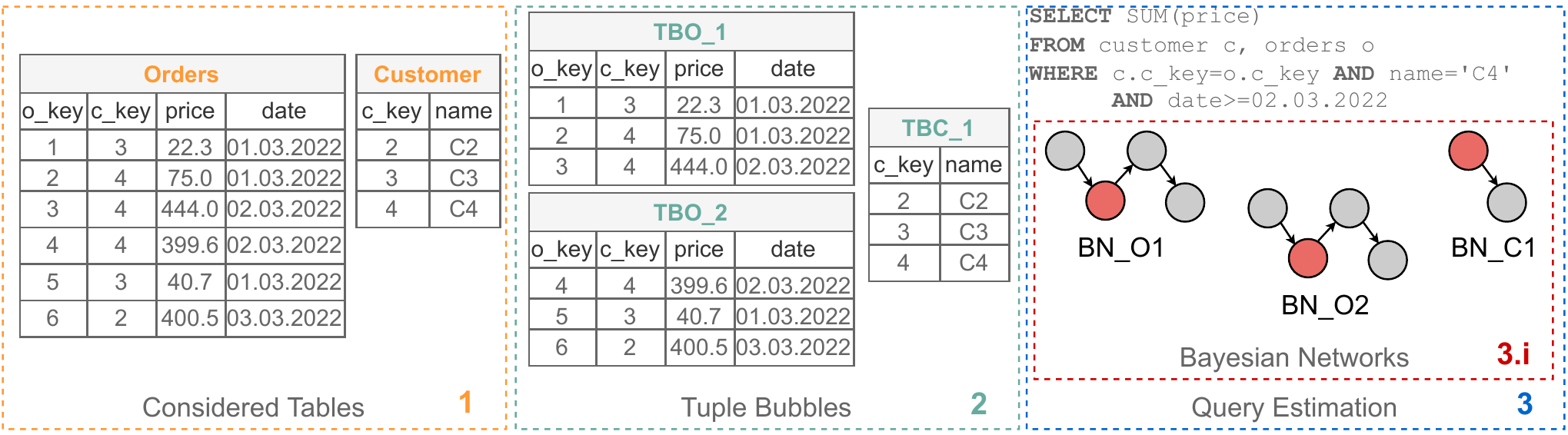}
  	\caption{Example dataset (1) and tuple bubbles (2); Bayesian networks for the tuple bubbles (3.i) and example  query (3)}
  	\label{fig:overall-approach-example}
\end{figure*}

\section{Bubbles Creation and Query Processing}
\label{sec:approach}

Bubbles can be derived from tables in various ways, ranging from plain horizontal partitioning, e.g., by primary key,  to more involved approaches such as identifying dependencies between tables and grouping similar tuples. The way bubbles are created naturally influences the subsequent statistics computation: Partitioning via ranges of primary keys is charming as it allows straightforward indexing of bubbles, while clustering tuples based on similar attributes allows  more concise and meaningful statistics. 

In this first work, we create bubbles by horizontal partitioning of the underlying tables. 
To avoid partitioning tables that do not have many records, the user needs to set the partitioning parameter $\theta$ indicating the minimal number of records that a relation should have and the parameter $k$ indicating the maximal number of bubbles that a table can be split into. 
An example of the bubbles creation is depicted in \figurename~\ref{fig:overall-approach-example}. For the considered tables, \textit{Orders} and \textit{Customer} (\figurename~\ref{fig:overall-approach-example} Part $1$), and the parameter $\theta=3$, we will form three tuple bubbles (two for the table \textit{Orders} and one for the table \textit{Customer}) as shown in  \figurename~\ref{fig:overall-approach-example} Part $2$. 

Depending on the availability of workload assumptions, bubbles can also be created for frequently occurring (sub) queries.
For well normalized databases, typically, there are many queries involving {\bf foreign-key joins}. We can make use of this such that bubbles can be created for joined partitions based on the foreign-key relations. For the \textit{Orders} and \textit{Customer} tables of Part $1$ in \figurename~\ref{fig:overall-approach-example}, once the horizontal partitioning is realized (Part $2$), the partitions for \textit{Orders} (TBO\_$1$ and TBO\_$2$) will be joined with the partition for \textit{Customer} (TBC\_$1$) adding the \textit{name} to the orders information. As a result, two bubbles will be formed. By doing this, we immediately cover  typical  joins between two relations and, thus,  can estimate query results more accurately.

\subsection{Bayesian Networks as Summaries}
\label{ssec:tuple_bubbles}

Once the records are structured into bubbles, we create representative but compact statistics. One simple approach is to use histograms. Although they will be easy to create, naturally they can answer only queries per columns and would not capture the dependencies between the columns. Furthermore, even more involved solutions that represent the data in chunks~\cite{liang2021combining} cannot answer queries that involve joins, which drastically affects their practical usability. In our work, we decided to use \textbf{Bayesian networks} which can precisely capture relationships between attributes, and leave the investigation of alternate solutions or hybrid variants over multiple different statistics for future work.

In the current approach, we instantiate one Bayesian network for each bubble
to represent the conditional dependencies between the enclosed tuples.
This means that for the three example bubbles in \figurename~\ref{fig:overall-approach-example} there are three Bayesian networks (Part $3.i$ \figurename~\ref{fig:overall-approach-example}). We envision that this creation will be parameterized and done in unison with the bubbles creation to give the user more freedom when deciding on their number and  content.

To identify a Bayesian network that closely matches the probability distribution between the attributes,
we make use of the Chow-Liu tree structure learning algorithm~\cite{chow1968approximating}. This method finds a Bayesian network by considering only dependencies between two attributes, forcing the network to have a tree like structure where the root is randomly chosen. 
Intuitively, the downside 
is that it can miss important attribute dependencies since it can not capture correlation between more than two attributes. 

To calculate the probability distribution for the nodes, we compute the conditional probability distribution for an attribute given the single parent attribute. The number of values for attribute $A_i$ will be $card(A_i)^{p+1}$ where $card(A_i)$ is the cardinality and $p$ is the number of parents which is one for our Bayesian network structure, i.e., $card(A_i)^2$ values need to be stored per distribution. Still, for attributes that have high value cardinalities a lot of values need to be stored. We employ the same approach as in related work~\cite{DBLP:journals/tlsdkcs/HalfordSM20,DBLP:journals/corr/abs-2012-14743,DBLP:conf/dasfaa/HalfordSM19} to compress the number of values. We store the exact probabilities for the $k$ most appearing values of the attribute. For the remaining values, we employ binning by grouping them into $b$ buckets where $b$ can be varied according to the memory requirements. The less appearing values are discretized into equal-sized buckets. Every bucket $b_i$ is identified by an integer id and together with the minimal and maximal value, the number of unique values for that range is stored. Using only the unique values for the buckets and the exact probability for the $k$ most frequent values is enough for producing accurate estimates. The algorithm for producing estimates from the Bayesian network is discussed in Section~\ref{sec:aqp_over_tuple_bubbles}.

\subsection{Query Processing}
\label{ssec:aqp_over_tuple_bubbles}

When a user issues a query, we access the appropriate bubbles and, with that, the Bayesian networks relevant to the query. For now, we propose that the query answer is estimated by creating \textit{as many substitute queries} as there are combinations of the bubbles for the relations of the query. For clarification, let us consider the example query from \figurename~\ref{fig:overall-approach-example}. There are three bubbles, two representing the \textit{Orders} table and one for the \textit{Customer} table. The query performs join between these relations and to answer it with our created statistics we would need two different queries, one joining TBO\_1 with TBC\_1 and the other TBO\_2 with TBC\_1. The estimate is then computed following the procedure in Section~\ref{ssec:estimating_join_queries}.

Evidently, if there are many tuple bubbles that should be considered for a given query, there will be many substitute queries that will need to be executed. In particular, if bubbles are created per relation, for a query that has $l$ relations that can be represented by $k_1, ..., k_l$ tuple bubbles, where $k_i$ is the number of bubbles for relation $R_i$, then $\prod_{i=1}^{l}k_i$ substitute queries would need to be estimated. Although the bubbles are represented through statistics and the computation will be much more efficient than considering all records from the complete relation, the needed transformations and preprocessing for the substitute queries will incur increased processing times. To mitigate the processing effects, instead of selecting all bubbles for a query we can pick $\sigma$ bubbles where $1 \le \sigma \le min(k_i)$, balancing between the estimation accuracy and  query execution time.

Let us consider that $\sigma$ is a small number such as $1$ and the tuple bubbles are chosen at random. In this scenario, it can happen that the chosen tuple bubbles do not hold any records that satisfy the query predicates or that there are no join results between them. Consequently, the query estimate will be exceptionally poor. To avoid these situations, we can store an \textit{additional compact index} for the bubble attributes. The index can then be utilized for guiding the selection. As a result, while incurring small additional memory occupancy we evade the unacceptable estimates.

To combine the estimates from the substitute queries and obtain the final query estimate $est(q)$ we employ the following formula for \texttt{SUM}, \texttt{COUNT}, and \texttt{AVG} queries. 
\begin{equation}
est(q) = \sum\limits_{i=1}^m est(qs_i) * weight_i
\label{eq:formula_combining_estimates_tuple_bubbles}
\end{equation}
The parameter $m$ is the number of substitute queries, $est(qs_i)$ is the estimate for the query $qs_i$. The parameter $weight_i$ will be $1$ for \texttt{SUM} and \texttt{COUNT} queries. For \texttt{AVG} queries $weight_i\!=\!N_{qs_i} / N$ where $N_{qs_i}$ is the number of results of the query $qs_i$ and $N$ is the total number of results from all relevant queries. A substitute query is relevant if it has at least one result. For \texttt{MIN} and \texttt{MAX} queries, the final estimate will be the minimal or maximal estimate from the substitute queries.

\section{Answering Queries over Bayesian Networks}
\label{sec:aqp_over_tuple_bubbles}

As we have seen, processing an aggregation query requires selecting the appropriate bubbles, aligning them in substitute queries, and aggregating the obtained results. Below, we discuss the estimation algorithms for a single Bayesian network, for queries involving selection predicates.
Then, we show how different Bayesian networks can be connected for answering join queries.  

\subsection{Inference over Single Bayesian Network}
\label{ssec:probability_inference_single_bn}

\begin{figure}[!t]
	\center
	\includegraphics[width=1.\linewidth]{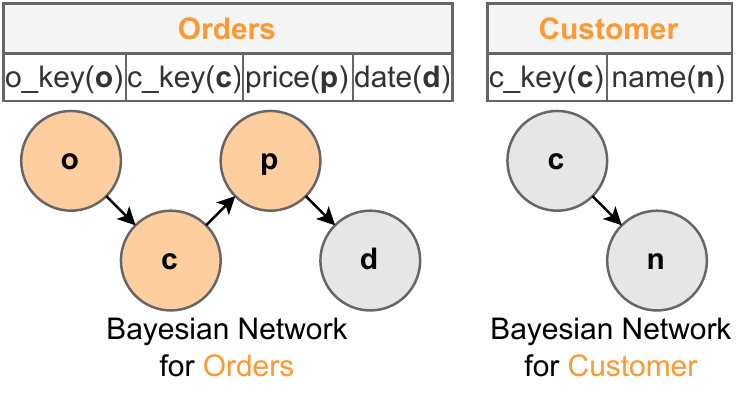}
  	\caption{Example Bayesian networks}
  	\label{fig:example-bayesian-networks}
\end{figure}

Let us consider that our example Bayesian network is created for the attributes of the relation \textit{Orders} as shown in \figurename~\ref{fig:example-bayesian-networks} and that we want to perform inference over it. In a Bayesian network, the process of estimating the probability of some subset of the attributes $A$ given some assignment of other attributes $E$ (evidence), i.e., $P(A|E)$, is called \textit{probability inference}. To compute this, we have to marginalize the joint probability distribution for all attributes that do not appear in $A$ and $E$. 
Since our created Bayesian networks will have a tree-like structure, we can utilize efficient algorithms that operate on trees. Following the related  work, we  make use of two inference algorithms. For exact inference, we use the \textit{variable elimination} algorithm~\cite{DBLP:conf/dasfaa/HalfordSM19,DBLP:journals/tlsdkcs/HalfordSM20,DBLP:journals/corr/abs-2012-14743}. Additionally, for faster but approximate estimation, we  utilize \textit{progressive sampling}~\cite{DBLP:journals/corr/abs-2012-14743,DBLP:journals/pvldb/YangLKWDCAHKS19}. 

\textbf{Variable elimination} is an efficient exact inference algorithm that can be applied to any kind of network~\cite{zhang1994simple}. The algorithm performs marginalization over the joint probability efficiently by moving the sum and product operations. It avoids repetitive computations and it operates on much smaller factors. For the example Bayesian network of the table \textit{Orders} (\figurename~\ref{fig:example-bayesian-networks}), to infer over the attribute \textit{price} (\textit{p}), the variable elimination algorithm can be applied as:
\begin{equation}
P(p) = \sum\limits_{o} P(o) \sum\limits_{c} P(c|o) * P(p|c) \sum\limits_{d} P(d|p) 
\label{eq:variable_elimination}
\end{equation}
By moving the summations inside, the algorithm reduces the computation by operating over much smaller relations. 
The algorithm can be further improved by eliminating the nodes from the network that are not required for obtaining a marginal distribution~\cite{DBLP:conf/dasfaa/HalfordSM19,DBLP:journals/tlsdkcs/HalfordSM20}. For our example, the relevant nodes when performing inference for the attribute \textit{price} are colored with orange in \figurename~\ref{fig:example-bayesian-networks}.

The main idea of \textbf{Progressive Sampling}  is to draw samples in iterations, progressively, such that the attributes will be traversed one at a time following a fixed ordering\cite{DBLP:journals/corr/abs-2012-14743,DBLP:journals/pvldb/YangLKWDCAHKS19}. In other words, the samples from the first attribute allow us to focus in a more relevant part of the second attribute and so on. 
To see how this works, consider a single Bayesian network that models the table \textit{Orders} as $P_O$. Let us assume that the user issues a \texttt{COUNT} query $q$ with arbitrary predicates $R(A_1), ..., R(A_4)$ that can represent any regions for the attributes of the table \textit{Orders}. The probability for the given query can then be expressed as:
\begin{equation}
\label{eq:probability_progressive_sampling}
P_O(q) = \prod_{i=1}^{4} P_O(A_i \in R(A_i) | Par(A_i) \in R(Par(A_i)))
\end{equation} 
where $Par(A_i)$ is the parent node of the attribute node $A_i$ and $R(Par(A_i))$ is the query region for that attribute. Consequently, to estimate the query probability, we need to estimate the conditional probability of the product for every attribute in Equation~\ref{eq:probability_progressive_sampling}. To efficiently compute this, the algorithm utilizes Monte Carlo approximation based on a sample $S$ of $R(Par(A_i))$ as $\dfrac{1}{S} \sum\limits_{s \in S} P_O(R(A_i)|s)$.

To execute the algorithm, first, the attributes are ordered as $A_1, ..., A_4$ where $A_1$ is the root of the Bayesian network. For the first attribute $A_1$, we can immediately obtain the probability $P_O(R(A_1))$ and use it to generate a sample $S_1$. 
For every next attribute, the samples for the parent have already been generated since the attributes follow a fixed ordering. Using the samples we can obtain a distribution that approximates $P_O(A_i|R(Par(A_i)))$. This distribution is used for the estimation and also for generating samples for the following attribute. This procedure is repeated for the remaining attributes. At the end, using the estimates, $P_O(q)$ can be obtained by taking their product. 

\textbf{Estimating Aggregate Queries:} 
As an output of the Bayesian network we obtain the probability of an assignment of one of the attributes ($A_i$), conditioned on one or multiple attributes (referred to as evidence $E$), i.e., $P(A_i | E)$. 
To create an estimate, the probabilities for the respective aggregation query need to be transfered. 

To estimate the result of a \texttt{COUNT} query, first, the algorithms traverse the nodes of the Bayesian network following their fixed order. At every step, it computes the conditional probability and multiplies it with the previously obtained probabilities. At the end, the query estimate is computed by taking into consideration the cardinality of the respective relation~\cite{DBLP:conf/dasfaa/HalfordSM19,DBLP:journals/tlsdkcs/HalfordSM20,DBLP:journals/corr/abs-2012-14743}. 
However, for other types of aggregation queries, i.e., \texttt{SUM}, \texttt{AVG}, \texttt{MIN}, and \texttt{MAX}, we need to make use of  per-value selectivities for the aggregation predicate to estimate the query result. Let us first consider \texttt{MIN} and \texttt{MAX} queries with an arbitrary number of selection predicates. The per value selectivities for the aggregation attribute can be easily interpreted as cardinalities following the above mentioned approach. Once we know the per value cardinalities of the attribute, we can consider only those values that appear at least once. We will return the minimum or maximum to answer the \texttt{MIN} or \texttt{MAX} query, respectively. For \texttt{SUM} queries, the cardinalities for the qualifying values will be summed up and returned as a result. For \texttt{AVG} queries, the summed up values will be divided with the cardinality of the same query.  

Since we employ binning for some attributes, we need to adjust the computations to consider ranges instead of single values. A range is identified by a minimal and a maximal value together with the number of distinct elements for the range. Consequently, we would consider the minimum or maximum when working with \texttt{MIN} or \texttt{MAX} queries, respectively. For \texttt{SUM} and \texttt{AVG} queries, the average value for the range is computed and used in the computation for the query. Although this preliminary approach for ranges provides appropriate results, it has potential for improvement.

\subsection{Estimating Join Queries}
\label{ssec:estimating_join_queries} 

Next, we will discuss the utilization of Bayesian networks for estimating queries involving joins. Let us consider that every relation in the database is represented by a separate tuple bubble. For our example, this means that there  are  two Bayesian networks, one for \textit{Orders} and another for \textit{Customer}, as depicted in \figurename~\ref{fig:example-bayesian-networks}. 
Let us assume that the user issues the same query as in part $3$ of \figurename~\ref{fig:overall-approach-example} but with \verb+COUNT+ instead of \verb+SUM+. 
To answer it, since there is one network per table, we have to resort to the join uniformity assumption. Consequently, the estimate will be 
$3/6 * 1/3 = 1/6$ 
instead of 
$2/6$ 
which means that the approach underestimates the result by $50$\%. Furthermore, if the query was asking for \verb+SUM+ it would be impossible to incorporate the customer name in the Bayesian network for \textit{Orders}, clearly leading to wrong results. 

To alleviate this problem, we propose to connect the Bayesian networks and \textit{incorporate the results from one Bayesian network into the others}. To accomplish this, when building the networks, we include all attributes for the relation, including the primary and foreign keys. Although the primary keys will not help us capture important correlations, they are necessary to enable the estimation over join queries. Capturing the primary-foreign key  relationships enables us to connect the Bayesian networks by using the shared nodes (attributes) between them. When having more than one Bayesian network, we  need to establish an appropriate ordering of them, making sure that every network can be connected through at least one attribute with the preceding and successive network. For the example query, since it includes the tables \textit{Customer} and \textit{Orders}, both Bayesian networks are selected. Then, we order them to form a \textit{chain following the primary-foreign key relation}, where the network that holds the aggregation attribute is last in the order. This is required because the probabilities of the aggregation attribute are needed to estimate the result. 

To connect the Bayesian networks, starting from the first network, we extract the probabilities for the attribute that is shared with the successive Bayesian network. When doing this, we take into consideration the relevant predicates from the query that can be applied on the network. The extracted probabilities are then used as evidence in the successive Bayesian network together with all the relevant query predicates for that network. The same procedure is repeated for all the networks until the last one. For the final Bayesian network, instead of extracting the probabilities for the aggregation attribute, we estimate the result according to the explanation in Section~\ref{ssec:probability_inference_single_bn}. 

For the example \texttt{COUNT} query, since \textit{price} is the aggregation attribute, the Bayesian network for \textit{Customer} will be the first one and the one for \textit{Orders} the second one in the order. The Bayesian networks can be connected through the attribute \textit{c\_key (c)}. For the first Bayesian network we will extract the probabilities for \textit{c} by applying the query predicate \verb+name='C4'+. This would return the value $4$. Then, for the Bayesian network of the \textit{Orders} table, we will consider the predicate \verb+c_key=4+ together with the query predicate \verb+date>=02.03.2022+. Rows $3$ and $4$ satisfy these predicates resulting in an accurate estimate for the query.

\begin{algorithm}[!t]
\centering
\caption{AQP over Tuple Bubbles}
\label{alg:aqp_over_tuple_bubbles}
\begin{algorithmic}[1]
\Function{EstimateResult}{$Q$, $TB$, $I_{TB}$, $\sigma$}
\State{$TB_{rel}=\{\}$; $TB_Q=[]$; $Q_{s}=[]$; $Q_{est}=[]$}
\State{$TB_{rel}=$ \textit{matchingBubbles}($Q.relations, TB$)}
\State{$TB_{Q}=$ \textit{extractTB}($Q$, $TB_{rel}$, $\sigma$, $I_{TB}$)}
\For{$tb_q$ \textbf{in} $TB_{Q}$}
	\State{$Q_{s}$.\textit{add}(\textit{replaceRelations}($Q.relations$, $tb_q$))}
\EndFor
\For{$i$ \textbf{in} $|Q_{s}|$}
	\State{$BN_{q_s}=TB_{Q}[i]$}
	\State{\textit{order}($BN_{q_s}$)}
	\State{$Q_{est}$.\textit{add}(\textit{estimate}($Q_{s}[i]$, $BN_{q_s}$))}
\EndFor
\State{\textbf{return }Equation~\ref{eq:formula_combining_estimates_tuple_bubbles}($Q_{est}$, $TB_{Q}$, $Q.type$)}
\EndFunction
\end{algorithmic}
\end{algorithm}

The complete approach for AQP over tuple bubbles is depicted in Algorithm~\ref{alg:aqp_over_tuple_bubbles}. As input, it receives the aggregation query $Q$, the tuple bubbles $TB$ represented through Bayesian networks, the index for the attributes of the tuple bubbles $I_{TB}$ and the parameter $\sigma$ restricting the number of tuple bubbles. Initially, the method identifies all bubbles that can be used for replacing the relations of $Q$ (Algorithm~\ref{alg:aqp_over_tuple_bubbles}, Line $3$). Subsequently, using the index $I_{TB}$, it selects $\sigma$ qualifying tuple bubbles aligning them to directly replace the relations of the query (Algorithm~\ref{alg:aqp_over_tuple_bubbles}, Lines $4$--$6$). For every substitute query, first the Bayesian networks will be extracted and ordered, and then the query will be estimated (Algorithm~\ref{alg:aqp_over_tuple_bubbles}, Lines $7$--$10$). The final estimate is produced by plugging in the substitute query estimates in Equation~\ref{eq:formula_combining_estimates_tuple_bubbles}.


\section{Challenges and Outlook}
\label{sec:challenges_future_outlook}

Despite the clear high-level idea of lifting query processing from tuples to bubbles, there are still several open research questions. 
First, the creation of the tuple bubbles needs to be thoroughly investigated. 
For very large datasets, joining tables upfront to create  bubbles for the result, might not be feasible. Horizontal partitioning can be problematic too, as it does not consider dependencies between the attributes of a table. If we follow a more suitable partitioning approach that can capture these dependencies we will be able to potentially create more meaningful and comprehensive bubbles. Additionally, if we take into consideration past query workloads for the creation of bubbles, we can create self-contained bubbles that store all the relevant records for a particular group of queries. 

Furthermore, as will be shown in our experiments, combining results from separate bubbles to form the final query estimate is difficult.
Specifically for \texttt{MIN} and \texttt{MAX} queries, for now, we can only estimate the result based on the minimal or maximal value in the considered bubbles and not over all the bubbles that satisfy the query. This limits our results since the approach heavily depends on the chosen bubbles. Additionally, combining results  for the substitute queries following traditional techniques produces estimates that although acceptable, can be further improved. The reason for this is the same as for the \texttt{MAX}/\texttt{MIN} queries although the results can be slightly improved considering basic statistics such as the bubble sizes. As one direction, we can consider storing compact index structures and representative thresholds for the individual bubbles and use them for better estimation. 

Although the Bayesian networks are able to capture the correlation between the attributes and with that provide results that in many scenarios outperform the competitors, other approaches for representing the tuple bubbles should be examined. In particular, when handling single tables, a structure such as PASS~\cite{liang2021combining} can accurately represent the data while occupying minimal space. Moreover, even between the bubbles or within a single bubble we can store different statistics that are better for answering different queries.


\section{Experiments}
\label{sec:experiments}

We have implemented the proposed approach in Python and conduct the experiments on a Linux machine with two $6$--core Intel Xeon E$5$-$2603$ v$4$ CPUs @$1.7$ GHz and $128$GB RAM. For the experiments, we consider three datasets and report on \textit{accuracy}, \textit{estimation time (i.e., latency)}, and \textit{required disk space} of our proposed approach and the considered competitors. We evaluate the following approaches:
\begin{enumerate}
  \item Our tuple bubbles approach, set up in four different flavors:
\begin{itemize}
  \item The bubbles represent the complete relations (\textbf{TB}) from the database. 
  \item The relations are horizontally partitioned into $k$ partitions and every partition represents a tuple bubble, where for estimation we use $1 \le i \le k$ bubbles for a relation (\textbf{TB\_i}). 
  \item Tables are joined based on the foreign-primary key relations and one tuple bubble (\textbf{TB\_J}) is created for every join result.
  \item Relations are horizontally partitioned into $k$ partitions and the partitions are joined based on the foreign-primary key relations. For each  join result, one bubble is created, and  $i$ bubbles from the same pair of relations are used for estimation (\textbf{TB\_J\_i}).
\end{itemize}

  \item PostgreSQL 14 serving as a baseline for producing exact answers.
  \item VerdictDB (\textbf{VDB})~\cite{park2018verdictdb} integrated in PostgreSQL and accessed through Java. We used scrambles of ratio $10$\% and $50$\%.
  \item Wander join (\textbf{WJ})~\cite{DBLP:journals/tods/LiWYZ19} integrated in PostgreSQL and accessed through Java. The approach can answer only \texttt{SUM} and \texttt{COUNT} queries, so, when evaluating, these are the only considered queries.
  \item \textbf{KD-PASS}~\cite{liang2021combining} using the authors' implementation in C++ with standard parameters as proposed by the authors. 
  \item The \textbf{AQP++}~\cite{peng2018aqp++} implementation of Liang et al.~\cite{liang2021combining} using the proposed parameters.
\end{enumerate} 
KD-PASS and AQP++ can only answer queries without joins, meaning that they are applicable only on single table datasets. 

\subsection{Datasets}
\label{ssec:datasets}
For the evaluation, we consider the following datasets.

\begin{itemize}

\item The international movies database benchmark (IMDB)~\cite{DBLP:journals/pvldb/LeisGMBK015} that consists of information about movies represented in $21$ tables. The dataset occupies $3.6$~GB when represented as CSV files. For the evaluation we use the job-light queries and create additional aggregation queries ($150$ in total) with $2$ to $5$ joins and $2$ to $5$ predicates. As an aggregation predicate we consider any of the numeric, integer or date attributes. 

\item The TPC-H benchmark that consists of $8$ tables. We generated $1$~GB of data and created $150$ aggregation queries involving $2$ to $5$ joins and from $2$ to $5$ predicates. The aggregation predicate can be any of the numeric, integer, or date attributes. 
\item Intel Wireless Dataset~\cite{intel_wireless_dataset} which is a single table dataset consisting of data from $54$ sensors. It contains of around $3$ million rows and has $8$ attributes where all attributes are continuous. For our evaluation, we considered all  attributes and we created $100$ aggregation queries with $2$ to $5$ predicates.

\end{itemize}

In our current version of the approach, we do not consider queries with group by or nested subqueries. 

\begin{table*}[!t]
    \centering
    \caption{Performance of the algorithms on the TPC-H dataset (The two values for the TB approaches are for PS and VE, respectively).}
    \centering
    \label{table:performance-tpch}
    \begin{tabular}{lcccccc}
        \toprule
        \multicolumn{1}{c}{} & \multicolumn{4}{c}{\textbf{Q-error (PS/VE)}} & \multicolumn{1}{c}{\textbf{Avg. Time}} & \multicolumn{1}{c}{\textbf{Memory}}\\
        \cmidrule(rl){2-5}  \cmidrule(l){6-6} \cmidrule(l){7-7}
          Approach                     & median        & 95th          & max             & avg    & ms            & MB\\
        \cmidrule(r){1-1} \cmidrule(rl){2-5} \cmidrule(l){6-6} \cmidrule(l){7-7}
          PostgreSQL                     & 1.0            & 1.0            & 1.0              & 1.0               & 1124.9       & 1399\\
\midrule
          \textbf{TB}              & 3.9 / 3.3     & 4620.0 / 4555.6 & 5.3*$10^4$ / 5.3*$10^4$    & 1064.0 / 1049.4 & \textbf{12.01} / \textbf{58.7} & \textbf{4.8}       \\
          \textbf{TB\_1}             & 2.3 / 2.29    & 4193.7 / 3810.5 & 5.3*$10^4$ / 3.8*$10^4$    & 1101.9 / 872.1 & \textbf{10.3} / \textbf{45} & \textbf{8.7}        \\
          \textbf{TB\_J}             & 1.2/ \textbf{1.018} & \textbf{2500.1 / 3100.6} & \textbf{9740.0 / 9740.0} & \textbf{220.1 / 241.5}   & \textbf{16.4} /160.4 & \textbf{10.8}    \\
          \textbf{TB\_J\_1}            & 2.02 / 2.006  & 2439.9 / 6477.7 & 3.1*$10^{4}$ / 1.1*$10^6$  & 861.4 / 1.1*$10^{4}$ & \textbf{16} / \textbf{150} & \textbf{17.3}  \\  
\midrule
          VDB 10\%                 & 1.18          & 2*$10^6$      & 1.1*$10^{10}$     & 7.8*$10^7$ & 96.1       & 147.5  \\
          VDB 50\%                 & 1.02          & 1.7*$10^6$    & 1.1*$10^{10}$     & 7.7*$10^7$ & 874.1      & 359     \\
          WJ                           & 1.1           & 4.9*$10^6$      & 9.7*$10^8$       & 3.2*$10^7$   & 143.3      & 702       \\      
        \bottomrule
    \end{tabular}
\end{table*}
\begin{table}[!t]
    \centering
    \caption{Performance of the algorithms on the IMDB dataset (only PS)}
    \centering
    \label{table:performance-imdb}
    \scalebox{0.85}{
    \begin{tabular}{lcccccc}
        \toprule
        \multicolumn{1}{c}{} & \multicolumn{4}{c}{\textbf{Q-error (PS)}} & \multicolumn{1}{c}{\textbf{Avg. Time}} & \multicolumn{1}{c}{\textbf{Memory}}\\
        \cmidrule(rl){2-5}  \cmidrule(l){6-6} \cmidrule(l){7-7}
          Approach                     & median        & 95th          & max             & avg    & ms            & MB\\
        \cmidrule(r){1-1} \cmidrule(rl){2-5} \cmidrule(l){6-6} \cmidrule(l){7-7}
          PostgreSQL                     & 1.0            & 1.0            & 1.0              & 1.0       & 1.6*$10^4$ & 7655  \\
\midrule
          \textbf{TB\_J}             & 2.2          & \textbf{106.0} & \textbf{1971.0} & \textbf{47.9} & \textbf{25} & \textbf{55} \\
          \textbf{TB\_J\_1}            & 6.2           & 2554.5        & 4.9*$10^4$      & 1870.3   & \textbf{19} & 89.3   \\
          \textbf{TB\_J\_3}            & 4.7           & 1116.2        & 7109.8      & 317.6  & \textbf{80} & 108.8   \\
\midrule
          VDB 10\%                 & \textbf{1.18} & 1971.0        & 1.9*$10^5$      & 3524.9   & 3255.6     & 395     \\
          VDB 50\%                 & \textbf{1.02} & 1224.0        & 9*$10^4$        & 1483.9   & 1.4*$10^4$ & 666     \\
          WJ                     & 2.44          & 1.7*$10^8$    & 1.7*$10^9$      & 6.5*$10^7$ & 108      & 5216   \\      
        \bottomrule
    \end{tabular}
    }
\end{table}

\subsection{Experimental Results}
\label{ssec:experimental_results}

We use $k=3$ and $\theta=500\,000$ for the horizontal partitioning of the tables, where $k$ is the maximal number of partitions and $\theta$ the minimal number of records for a table to qualify for partitioning. For the binning in the Bayesian networks, we use between $60$ to $200$ buckets. For the categorical attributes, we directly store information about the $40$ to $100$ most common values and put the remaining values in buckets. For our approach, we evaluate both proposed estimation algorithms from Section~\ref{ssec:probability_inference_single_bn}, progressive sampling (\textbf{PS}) and variable elimination (\textbf{VE}). For PS, we always use $1000$ samples.  For the accuracy evaluation we used the q-error which is the relation between the true result and the estimate, i.e., 
$q\_err = max(true(q)/est(q),est(q)/true(q))$. 

First, we report on results for TPC-H (Table~\ref{table:performance-tpch}) and IMDB (Table~\ref{table:performance-imdb}). For these datasets, we did not consider KD-PASS and AQP++ since they cannot answer queries involving joins. When analyzing the results for the TPC-H dataset, it is evident that the opponents have unacceptably high average and maximal errors. Additionally, when considering the execution time and memory, it is evident that for all scenarios, our approach drastically outperforms them. Concerning the different versions of our approach, building bubbles per table (TB) yields the worse results. However, if we partition the tables horizontally and consider only one partition per table for answering the queries (TB\_1) we achieve estimates with acceptable quality, estimation time and memory. Although our results are superior to those of the competitors, still there is room for improvement. To better capture the join queries involving more than two tables, we let the bubbles represent complete foreign-primary key joins (TB\_J). This version produces the best estimates overall, where the VE algorithm has the best median accuracy. Considering bubbles where the tables are first partitioned then the partitions are joined where one bubble per join is used for estimation (TB\_J\_1), it is observable that the accuracy is significantly worse for the higher percentiles, indicating that the connection of the results can be further improved.

Guided by the results for the previous dataset, for IMDB (Table~\ref{table:performance-imdb}), we consider only the PS algorithm and bubbles that represent foreign-primary key joins (TB\_J) and partitions that are joined (TB\_J\_i). The TB\_J approach again drastically outperforms the competitors in terms of accuracy, execution time and memory. However, for the partitioned variant, using only one bubble per join (TB\_J\_1), will not provide appropriate  results in terms of accuracy. The main reason for this are the \texttt{MIN} and \texttt{MAX} queries since to answer them we always return the minimum or maximum for the investigated bubbles not considering the data from the other bubbles. However, using three bubbles per join already provides results of acceptable quality in satisfactory time while requiring drastically less disk space than the competitors. Although the optimal solution for creating the final query estimate is an open problem, the  initial results support the conclusions that moving the processing from per tuple to per bubble can be highly beneficial.

\begin{table*}[!t]
    \centering
    \caption{Performance of the algorithms on the Intel dataset (The two values for the TB approaches are for PS and VE, respectively).}
    \centering
    \label{table:performance-intel}
    \begin{tabular}{lcccccc}
        \toprule
        \multicolumn{1}{c}{} & \multicolumn{4}{c}{\textbf{Q-error}} & \multicolumn{1}{c}{\textbf{Avg. Time}} & \multicolumn{1}{c}{\textbf{Memory}}\\
        \cmidrule(rl){2-5}  \cmidrule(l){6-6} \cmidrule(l){7-7}
          Approach                     & median        & 95th          & max             & avg    & ms            &MB\\
        \cmidrule(r){1-1} \cmidrule(rl){2-5} \cmidrule(l){6-6} \cmidrule(l){7-7}
          PostgreSQL                     & 1.0            & 1.0            & 1.0              & 1.0               & 153.5       & 201\\
\midrule
          \textbf{TB}              & 1.24 / 1.1    & 44.6 / 43.7  & 2720.0.0 / 2964.0 & 141.1 / 109.9  & \textbf{10.7} / 136.1 & \textbf{0.9}    \\
          \textbf{TB\_1}             & 1.57 / 1.31   & 1240.3 / 2789.8 & 7.7*$10^5$ &  1*$10^4$ / 2.9*$10^4$ & \textbf{9.5} / 114.5 & \textbf{0.89}  \\
          \textbf{TB\_2}             & 1.32 / 1.17   & 1331.3 / 1230.8 & 7.7*$10^5$ &  1*$10^4$ / 1*$10^4$ & \textbf{20} / 230.3 & \textbf{1.8}  \\
          \textbf{TB\_3}             & 1.17 / 1.12   & 1800.1 / 1179.0 & 7.7*$10^5$ &  1*$10^4$ / 8681.7 & \textbf{26} / 578.6 & \textbf{2.7}  \\
\midrule
          VDB 10\%                 & \textbf{1.003}& 79.0          & 5.9*$10^4$      & 1265.3       & 55.2       & 22     \\
          VDB 50\%                 & \textbf{1.001}& 129.0         & 5.9*$10^4$      & 1508.1       & 135.6      & 55     \\
          WJ                     & \textbf{1.005}&\textbf{1.9}   & \textbf{25.8}   & \textbf{2.03}& 155.3      & 90     \\ 
          KD-PASS                   & 1.04       & 690.8         & 2.2*$10^4$      & 318.3    & 47    & \textbf{0.1}       \\
          AQP++                  &  1.042        & 809.9         & 2.2*$10^4$      & 327.4    & 52    & \textbf{0.1}       \\
        \bottomrule
    \end{tabular}
\end{table*}

The results for the Intel dataset are shown in Table~\ref{table:performance-intel}. Since the queries do not involve joins, we can consider KD-PASS and AQP++ in the evaluation. When  tuple bubbles represent the $3$ partitions of the table, we evaluate the approach when using $1$, $2$, or $3$ of the bubbles for estimation, TB\_i respectively. We can see that,  although our approach where the bubbles represent the complete table (TB) has comparable median accuracy to the best competitor WJ, the estimates for the higher percentiles and the average are unsatisfactory. This is not unexpected since our approach uses buckets to represent continuous attributes and all the attributes in this dataset are of this form. Naturally, this  negatively affects the estimation. However, our approach requires drastically less disk space than most competitors. Additionally, the estimation using the PS algorithm can produce results in the best time. When analyzing the results using bubbles that represent partitions of the table, we can observe that the accuracy for the higher extremes becomes worse. As the number of bubbles used for estimation increases, so does the median accuracy which for using all the created bubbles (TB\_3) even exceeds the accuracy of the TB approach. Although the required disk space and execution time increase, still they are better than most of the competitors. 

Based on this results, moving the processing from tuples to tuple bubbles is beneficial. However, connecting the results is still an open issue since clearly, heedlessly connecting estimates from the bubbles is not the best approach when considering the higher percentiles. When working with single tables, PASS can be considered over Bayesian networks for representing the statistics since the queries will not have joins.


\section{Conclusion}
\label{sec:conclusion}
We addressed the problem of performing AQP by proposing to lift query processing from per tuples to statistical descriptions of groups of tuples, 
coined tuple bubbles. To accomplish a concise but accurate description, we proposed the usage of Bayesian networks and investigated their suitability to compactly represent the tuple bubbles and their ability to answer queries involving an arbitrary number of equality joins and equality or range predicates. We further detailed on encountered challenges  before validating our ideas through a preliminary evaluation. The proposed approach for moving the processing from tuples to bubbles can be highly beneficial in a distributed environment when handling data that cannot be efficiently processed on a single machine or if data shipping is prohibitively expensive. Investigating such an environment should be complementary to the analysis of possible partitioning algorithms and  methods for combining the results. 

\balance
\bibliographystyle{plain}
\bibliography{main}

\end{document}